\documentclass[lettersize,journal]{IEEEtran}
\usepackage{amsmath,amsfonts}
\usepackage{algorithmic}
\usepackage{algorithm}
\usepackage{array}
\usepackage[caption=false,font=normalsize,labelfont=sf,textfont=sf]{subfig}
\usepackage{textcomp}
\usepackage{stfloats}
\usepackage{url}
\usepackage{verbatim}
\usepackage{graphicx}
\usepackage{cite}
\usepackage[dvipsnames]{xcolor}
\usepackage{microtype}
\usepackage{colortbl}
\usepackage{soul,xcolor}
\sethlcolor{red!30}
\usepackage{xcolor}
\usepackage{array}
\usepackage{graphicx}  
\usepackage{multirow}  

\setul{0.1ex}{0.2ex}

\begin{document}

\setstcolor{red}

\title{A Comprehensive Design Framework for UE-side and BS-Side RIS Deployments}

\author{Mahmoud~Raeisi,~\IEEEmembership{Student Member,~IEEE},
        Aymen~Khaleel,~\IEEEmembership{Member,~IEEE},
        Mehmet C.~Ilter,~\IEEEmembership{Senior Member,~IEEE},
        Majid~Gerami,~\IEEEmembership{Member,~IEEE},
        and
        Ertugrul~Basar,~\IEEEmembership{Fellow,~IEEE}

        \vspace{-2em}
        
\thanks{M. Raeisi, A. Khaleel, and E. Basar are with the Communications Research and Innovation Laboratory (CoreLab), Department of Electrical and Electronics Engineering, Koç University, Sariyer, Istanbul 34450, Turkey. (e-mail: mraeisi19@ku.edu.tr; akhaleel18@ku.edu.tr; ebasar@ku.edu.tr)}

\thanks{Mehmet C. Ilter was with Huawei Lund Research Center during this work and is currently with the Department of Electrical Engineering, Tampere University, Finland (mehmet.ilter@tuni.fi).}

\thanks{M. Gerami is with the Lund Research Center, Huawei Technologies Sweden AB, Sweden. (e-mail: majid.gerami@huawei.com)}
}




\maketitle

\begin{abstract}
Integrating reconfigurable intelligent surfaces (RISs) in emerging communication systems is a fast-growing research field that has recently earned much attention. While implementing RISs near the base station (BS), i.e., BS-side RIS, or user equipment (UE), i.e., UE-side RIS, exhibits optimum performance, understanding the differences between these two deployments in terms of the system design perspective needs to be clarified. Critical design parameters, such as RIS size, phase shift adjustment, control link, and element type (passive/active), require greater clarity across these scenarios. Overlooking the intricacies of such critical design parameters in light of 6G demands endangers practical implementation, widening the gap between theoretical insights and practical applications. In this regard, our study investigates the impact of each RIS deployment strategy on the anticipated 6G requirements and offers tailored RIS design recommendations to fulfill these forward-looking requirements. Through this, we clarify the practical distinctions and propose a comprehensive framework for differentiating between BS-side and UE-side RIS scenarios in terms of their design parameters. Highlighting the unique needs of each and the potential challenges ahead, we aim to fuse the theoretical underpinnings of RIS with tangible implementation considerations, propelling progress in both the academic sphere and the industry.
\end{abstract}



\section{Introduction} 

Reconfigurable intelligent surfaces (RISs) represent an emerging technology with significant potential for sixth-generation (6G) communication networks. RISs offer the capability to actively mitigate the detrimental effects of the propagation environment, such as multipath fading \cite{basar2019wireless}. In other words, RISs pave the way for the realization of \textit{smart radio environments}, where the environment itself becomes an active participant in wireless transmission by dynamically managing signal reflection/refraction across different locations \cite{basar2019wireless}.

{One of the most compelling applications of RISs is their deployment in close proximity to either the base station (BS) or the user equipment (UE). This strategic placement, known as BS-side and UE-side RIS \cite{9745477}, offers distinct advantages tailored to specific network requirements. However, what exactly do these terms mean, and why are they critical for future networks?}

{BS-side RIS refers to a scenario where the RIS is positioned near the BS, {enabling a reliable control link as the wireless service provider can efficiently establish and maintain control links between the BS and its RIS units. This setup simplifies the configuration process by leveraging the BS’s advanced hardware capabilities.} On the other hand, UE-side RIS involves placing the RIS near the UE, which can significantly enhance signal quality in challenging environments, such as areas with poor coverage or high interference.} {However, establishing reliable control links between BS and UE-side RISs can be challenging as these RIS units may be distant from the BS without a strong line-of-sight (LOS) connection. Additionally, a high number of UE-side RIS units could impose substantial control signaling overhead on the BS. Hence, the wireless service provider cannot efficiently control UE-side RISs.}

{By understanding the unique benefits and challenges associated with each deployment strategy, we can better tailor RIS designs to meet the specific demands of 6G networks. The following sections explore how these strategies can be optimized to achieve the best performance in various scenarios.}

\begin{figure*}
    \centering
    \includegraphics[width = \textwidth]{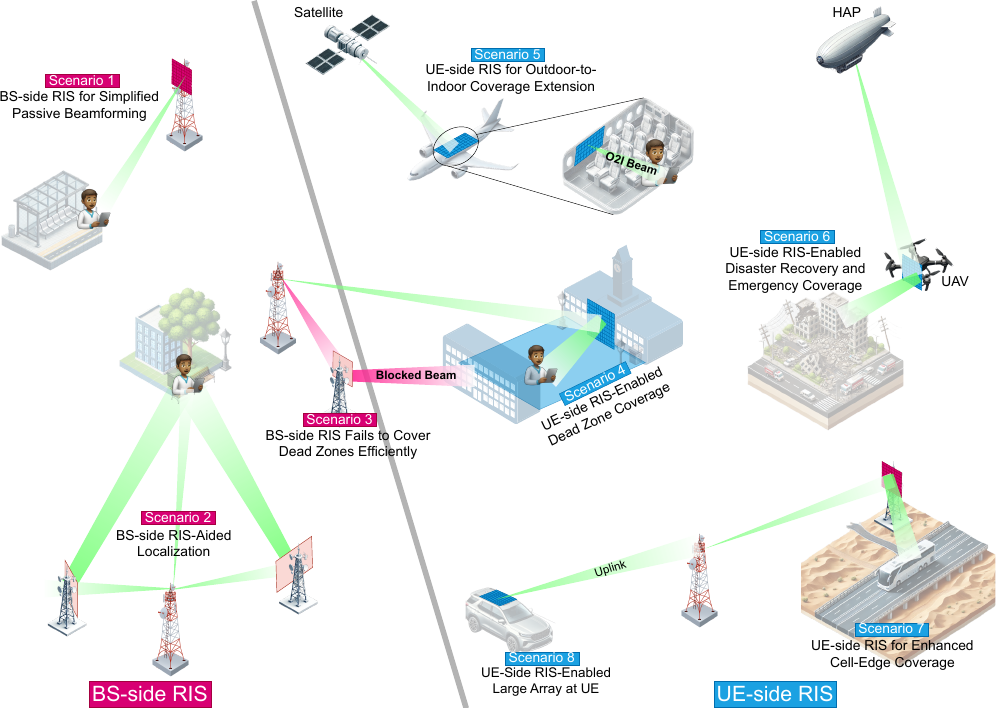}
    \caption{{BS-side and UE-side RIS deployments within a generic RIS-assisted wireless network.}}
    \label{fig:General-Applications}
\end{figure*}

Fig. \ref{fig:General-Applications} showcases diverse RIS applications {tailored to address specific challenges in wireless communications systems}. 
{As a \textit{BS-side} application, RISs can simplify the hardware design of BS by using passive elements instead of active power-hungry components, reducing both power consumption and overall system cost {(scenario 1 in Fig. \ref{fig:General-Applications})} \cite{9162097, 10144102}.}
{On the other hand, the \textit{UE-side RIS} shines in scenarios where user mobility and varying coverage conditions (like dead zones and cell edges {which may encompass a various range of conditions, such as Scenarios 4–7 illustrated in Fig. \ref{fig:General-Applications}}) present significant challenges.} This challenge is particularly pronounced in high-frequency communication systems like millimeter-wave (mmWave) and terahertz bands. Traditionally, covering entire environments requires deploying additional BSs or relays, but this can be costly and inefficient. RISs offer a cost/energy-efficient solution as a programmable reflector network within the environment, addressing the coverage dilemma \cite{raeisi2023plug, 9852389}. 
{The UE-side} RISs can {also} realize the implementation of large arrays in the UE in a more compact and cost-efficient manner {(Scenario 8 in Fig. \ref{fig:General-Applications})} \cite{9598898}. 

{While BS/UE-side RIS deployments are particularly effective in the scenarios mentioned above, their applications extend far beyond these examples. These deployments are poised to play a crucial role in a wide range of emerging 6G technologies, including non-terrestrial networks {(Scenarios 5 and 6)}, localization and sensing {(Scenario 2)}, vehicular networks {(Scenarios 5, 7, and 8)}, and more.}

Although current literature considers both BS-side and UE-side RIS system designs, it is still unclear how their design parameters change from one to another. In particular, design parameters related to the RIS, such as its size, phase shift adjustment, control link, placement and deployment, and elements type (passive/active), are yet to be differentiated among the two scenarios (BS-side and UE-side). Overlooking these parameters' differences while designing the RIS-assisted system affects their implementation feasibility, leading to a big gap between theoretical research and industry. For instance, when we examine the control link as an essential design parameter, it becomes apparent that the BS-side RIS system benefits from a more reliable control link. This advantage arises from the static positioning of both the BS and the RIS, enabling the utilization of a robust LOS wireless backhaul link or even a fiber optic cable. In contrast, ensuring a similarly robust control link in the UE-side RIS system is challenging, primarily attributed to resource constraints and UE's mobility. Extending this discussion to the other design parameters can reveal critical issues between the theoretical assumptions and implementation restrictions, which motivates this study.

In this article, we investigate the RIS-related design parameters for the BS/UE-side scenarios, shedding light on the critical differences that need to be considered by system designers. Specifically, by considering these parameters individually, we discuss how each scenario has different requirements/settings for that parameter. Accordingly, we provide a practical guideline {to} distinguish between these two scenarios from the system design and implementation perspectives.

\section{Key Aspects of RIS Deployment Framework}
\label{Sec: RIS placement}

{Optimizing RIS deployment strategies is crucial for enhancing performance across various communication scenarios. Understanding the differences between BS-side and UE-side implementations is essential to address practical deployment considerations. This section explores these deployment strategies and highlights the key design parameters influencing their effectiveness.}

\subsection{{Deployment Strategies: BS-side and UE-side RIS}}

{An RIS deployment is considered BS-side when the RIS is located near the BS. {In this case,} the {wireless service provider} establishes a backhaul control link {between the BS and} RIS. When phase shift reconfiguration is required, the BS computes the optimal phase profile based on the current channel state information (CSI) and communicates this profile to the RIS control unit.}

{{Conversely}, in a UE-side RIS deployment, the RIS is positioned near the UE. {As discussed earlier, in this case, the practical issues like lack of reliable control link and heavy signaling overhead preventing BS to control the RIS. Hence, in this case, UE} is responsible for optimizing and controlling the RIS. {Additionally}, the RIS design should prioritize simplicity and energy efficiency, reflecting the UE's limited computational resources and mobility constraints.}

\begin{figure}
    \centering
    \includegraphics[scale = 0.35]{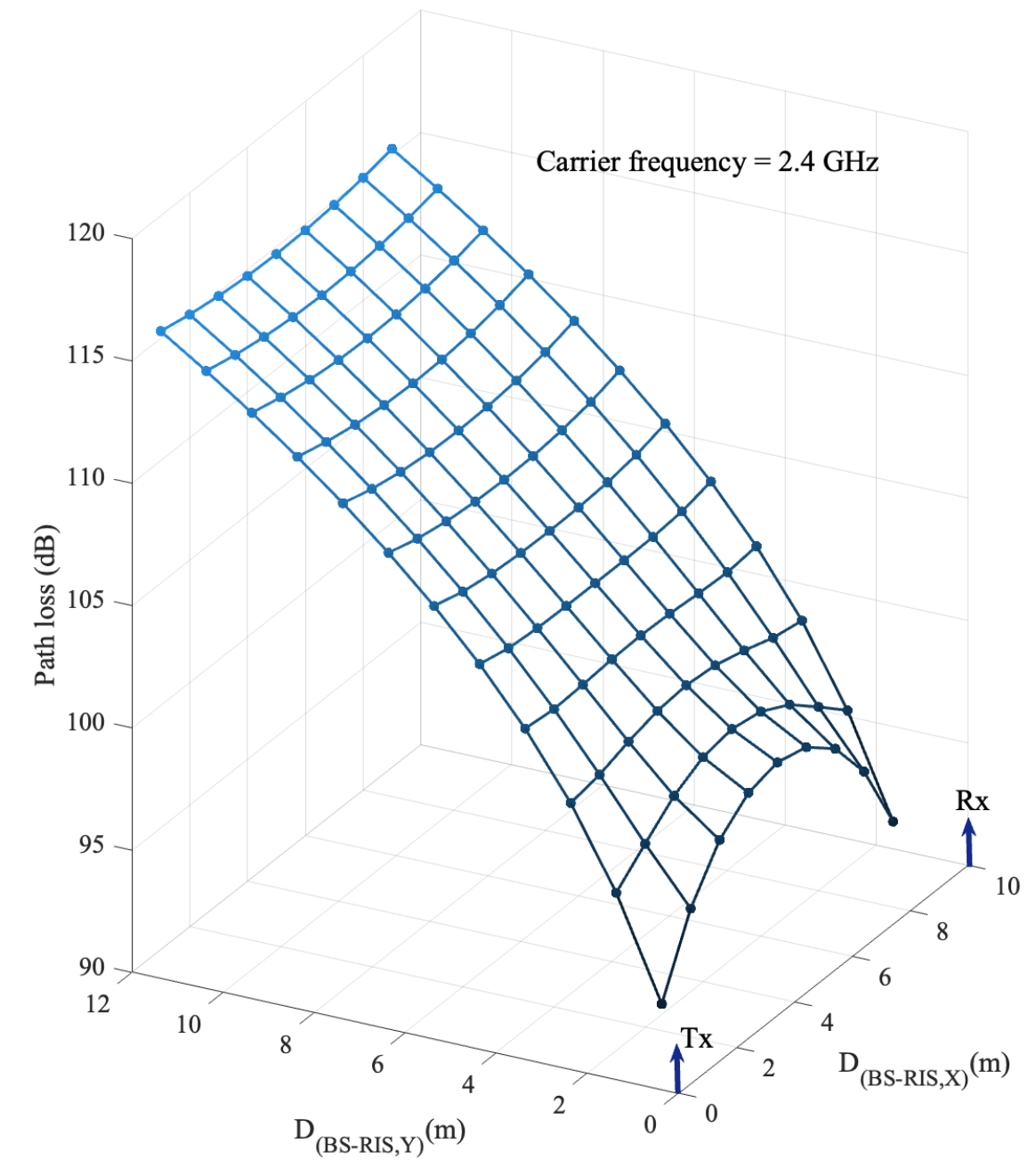}
    \caption{Two-way path loss in an RIS-assisted system over different RIS locations.}
    \label{fig:SNR vs RIS locations}
\end{figure}

In principle, the potential for RIS placement is virtually limitless, with the possibility of deploying these surfaces across diverse locations within the environment; however, practical limitations narrow down these possibilities. 
The primary constraint related to deploying RIS lies in the significant path loss associated with the signals imping and reflecting from its surface. 
Fig. \ref{fig:SNR vs RIS locations} suggests that integrating the RIS within the transceivers {minimizes free-space multiplicative path loss and} offers optimal placement; nevertheless, {it does not count the effect of other factors like, shadowing, large/small scale fading, space limitation, UE movement, and environmental obstacles.} {Such factors result in optimal RIS placement separated from the BS/UE.} This leads to classifying each deployment strategy into two categories. {When the RIS is closely integrated with the BS or UE, it is termed a BS/UE-side integrated RIS. If the RIS is deployed separately and not physically attached to the BS/UE, it is called a BS/UE-side separated RIS.}

Note that in Fig. \ref{fig:SNR vs RIS locations}, {the BS and UE are positioned at the coordinates $(0,0)$ and 
$(10,0)$, respectively. The RIS is deployed within the same 2D plane, with the projection of each simulated point onto the 
$XY$ plane representing the location of the RIS.} $D_{\textrm{(BS-RIS,i)}}$ represents the distance between BS and RIS along $i \in \{ X, Y \}$ axis while $D_{\textrm{BS-UE}}$ is the distance between BS and UE. Although {path loss generally increases as the RIS is deployed farther from the terminals, exceptions do exist.} Some cases necessitate us to consider $D_{\textrm{(BS-RIS,Y)}}$ constant. In such cases, when $D_{\textrm{(BS-RIS,Y)}} \geq D_{\textrm{BS-UE}}/2$, the best placing for RIS is $D_{\textrm{(BS-RIS,X)}} = D_{\textrm{BS-UE}}/2$, as shown in Fig. \ref{fig:SNR vs RIS locations} and discussed in \cite{10172310}. However, this is an exceptional case that does not happen frequently.

\subsection{{Key Design Parameters}}

{The performance and efficiency of RIS deployments are determined by several key design parameters that must be carefully balanced to achieve optimal results. These parameters—such as reliable control link, RIS size, phase shift adjustment, and element type—are not independent; rather, they are interconnected and jointly influence critical aspects like channel estimation overhead, resource allocation, and system complexity. A comprehensive understanding of these interrelations is crucial for developing a robust deployment strategy, whether in BS-side or UE-side scenarios.}

\begin{enumerate}
    \item {\textit{Control Link:} Establishing a reliable control link is essential for dynamically adjusting the phase shifts of the RIS elements based on CSI. Depending on the deployment, this link can be either wired or wireless. In BS-side RIS, the proximity to the BS allows for more stable and reliable control link, often facilitated by a LOS connection or even a wired backhaul. This leads to reduced channel estimation overhead since the statistical CSI (S-CSI) dominates over instantaneous CSI (I-CSI), allowing for a less frequent estimation process. Conversely, in UE-side RIS, the wireless control link faces greater challenges due to the mobility of the UE, the increased likelihood of obstacles, and the potential for a more dynamic channel, all of which require more frequent CSI updates.}

    \item {\textit{Phase Shift Adjustment:} The flexibility and resolution of RIS phase shift adjustments are crucial for performance. BS-side RIS levera ges the BS's advanced capabilities for high-resolution/continuous adjustments to optimize performance. In contrast, UE-side RIS relies on simpler, low-complexity configurations suited to the UE's limited resources to balance precision and efficiency.}

    \item {\textit{RIS Size:} The physical dimensions of the RIS directly affect its beamforming capabilities and overall effectiveness. Larger RIS panels provide greater beamforming gain and are more feasible in BS-side deployments, where space, stable mounting points, and complex hardware for resource allocation are available. In contrast, UE-side deployments face size constraints and limited resources, restricting the use of large-scale RIS and requiring more compact designs that still ensure sufficient performance in mobile contexts.}

    \item {\textit{Element Type:} The choice between passive and active elements in RIS design depends on deployment strategy and performance needs. Passive elements, being more energy-efficient, are preferred in BS-side RIS due to the available space for larger implementations. In contrast, UE-side RIS may require active elements to amplify signal strength, though this comes with higher complexity and power consumption.}
\end{enumerate}

\section{BS-Side RIS Deployment}

This section presents an overview of the BS-side RIS deployment, including its importance and potential applications.
 
In practice, the BS maintains a fixed location, allowing for strategic placement/orientation of the RIS to establish a LOS link with the BS. Furthermore, the proximity and immobility of the BS and the RIS in this design simplify channel estimation and enable a reliable control link from the BS to the RIS, as discussed in Section \ref{Sec: RIS placement}. Besides, owing to the more resources and sophisticated hardware available on its side (compared to the UE side), the BS can support advanced functionalities such as cascaded channel estimation/decoupling and joint active and passive beamforming optimization with high-resolution phase shifts to optimally configure the RIS. Additionally, BSs are typically positioned at elevated locations with plentiful space in the surrounding environment, making deploying a large RIS surface possible. A large-size RIS provides significant beamforming/multiplexing/array gain, reducing the overall path loss while maintaining cost-effective and energy-efficient hardware. 

To illustrate the potential of BS-side RIS applications, we present two examples highlighting its advantages in alignment with the anticipated 6G requirements in terms of energy/spectral efficiency enhancement.

\begin{enumerate}
    \item \textit{Enhanced energy efficiency}:
    In \cite{9162097}, a BS-side RIS design is proposed to reduce the required number of RF chains at the BS side. This approach involves implementing the RIS to redesign the classical Alamouti's scheme with a single RF chain instead of the traditional two, reducing cost and power consumption on the transmitter side. The advantage of such an approach in designing the BS is more evident in the generalized scenario of Alamouti's scheme: a large-scale RIS-assisted space-time block code (STBC) MIMO system, {leading to considerable power saving and enhanced energy efficiency.}

    \item \textit{Enhanced spectral efficiency}: Integrating passive RISs directly into the BS structure can facilitate passive beamforming/multiplexing cost-effectively \cite{10144102}, as {illustrated in the Scenario 1} in Fig. \ref{fig:General-Applications}. This novel design lets us directly reconfigure the signals incoming/outgoing from the BS without encountering substantial path loss due to the extremely short distance between BS and RIS. In this regard, the authors demonstrated that deploying an integrated RIS with the BS yields superior performance in terms of spectral efficiency compared to the conventional system without RIS.
\end{enumerate}

\section{UE-Side RIS Deployment}\label{Sec:UE-side RIS}

Another practical approach for RIS deployment is to place it near a single UE or a cluster of UEs where the {UE-RIS channel is more stable than BS-RIS, and UE has control over the RIS.} 

Due to UEs' mobility, they may end up in regions with weak/no network coverage. The conventional solution of adding more BSs/relays to address this issue can be practically costly and energy inefficient \cite{basar2019wireless}. Specifically, this challenge becomes more prominent when higher frequencies (like mmWave and terahertz bands) are adopted in future 6G networks, causing manifold dead zones due to the vulnerability to signal blockages. 
Since a BS-side RIS may encounter similar obstacles as the BS alone, deploying the RIS on the UE side is a more favorable solution to address the coverage issue. Thus, the RIS can establish an alternative link to the main blocked one by providing reflection paths reaching the dead zone. 
By strategically positioning each RIS in UE proximity, a reliable LOS link between the RIS and BS/UE can be guaranteed \cite{raeisi2023plug}. 
A relatively more reliable control link (compared to the BS-RIS link) enables the UE to control the RIS independently from the BS.

Here, we present two examples from the current literature to provide concrete context to the concept of UE-side RIS and its impact on the 6G requirements.

\begin{enumerate}
    \item \textit{Enhanced reliability:} As discussed in \cite{basar2021reconfigurable}, a primary use-case for RIS is extending the coverage to the cell edge and dead zone when the direct link between the BS and UE is weak or blocked, as illustrated {in Scenarios 4, 6, and 7} in Fig. \ref{fig:General-Applications}. Another attractive scenario is the coverage extension in an outdoor-to-indoor (O2I) setup where the BS is implemented outdoors while the UE is located indoors and receives weak/no signals \cite{10056847}, {as depicted {in the Scenario 5} in Fig. \ref{fig:General-Applications}}. 
    
    \item \textit{Enhanced user-experienced data rate:} 
    Enabling a large-scale antenna array at the UE has been seen as unfeasible due to size, energy consumption, and cost limitations \cite{9598898}. Nevertheless, the recent study of \cite{9598898} has introduced a UE-side RIS design that leverages multi-layer RIS surfaces to achieve a system performance similar to that obtained using a large-scale antenna array. Consequently, an amplified beamforming/array gain can be realized, raising the user-experienced data rate.
\end{enumerate}

\section{Key Considerations in RIS Tuning}

In order to achieve optimal performance in an RIS-assisted system, a dynamic RIS configuration based on I-CSI is required. The responsible terminal calculates the optimal phase shift profile and conveys this information to the RIS's control unit.
Inefficient control and configuration can negatively affect the system's overall performance by limiting the functionality of the RIS. 
{In what follows, we explore the key considerations involved in RIS tuning, focusing specifically on the practical aspects of control link management and phase shift configuration for both BS-side and UE-side scenarios.}

\subsection{Control Link}
In order to guarantee optimal {RIS tuning}, a robust control link with sufficient capacity to transmit control information from the responsible terminal to the RIS needs to be established.
Based on the operating scenario and RIS deployment strategy, selecting a control link medium can encompass various options, such as fiber-optic cables, metal-based wiring, or wireless links.

For BS-side RIS deployments, where both the BS and RIS have fixed locations, fiber-optic cabling can be the optimal choice for the control link regarding capacity and reliability. Nevertheless, a wireless link can be an alternative option for more flexible RIS placement/orientation to achieve {optimum tuning}.

In contrast, when deploying the RIS on the UE side, the cabling/wiring solution is often impractical due to UE mobility. Establishing a dedicated wireless control link is the most feasible solution. The only exception is when the RIS is integrated within the UE, as the stationary positions of the UE and RIS relative to each other make metal-based wiring a viable option for creating a dedicated control link \cite{9598898}.

\begin{figure*}
    \centering
    \includegraphics[width=\textwidth]{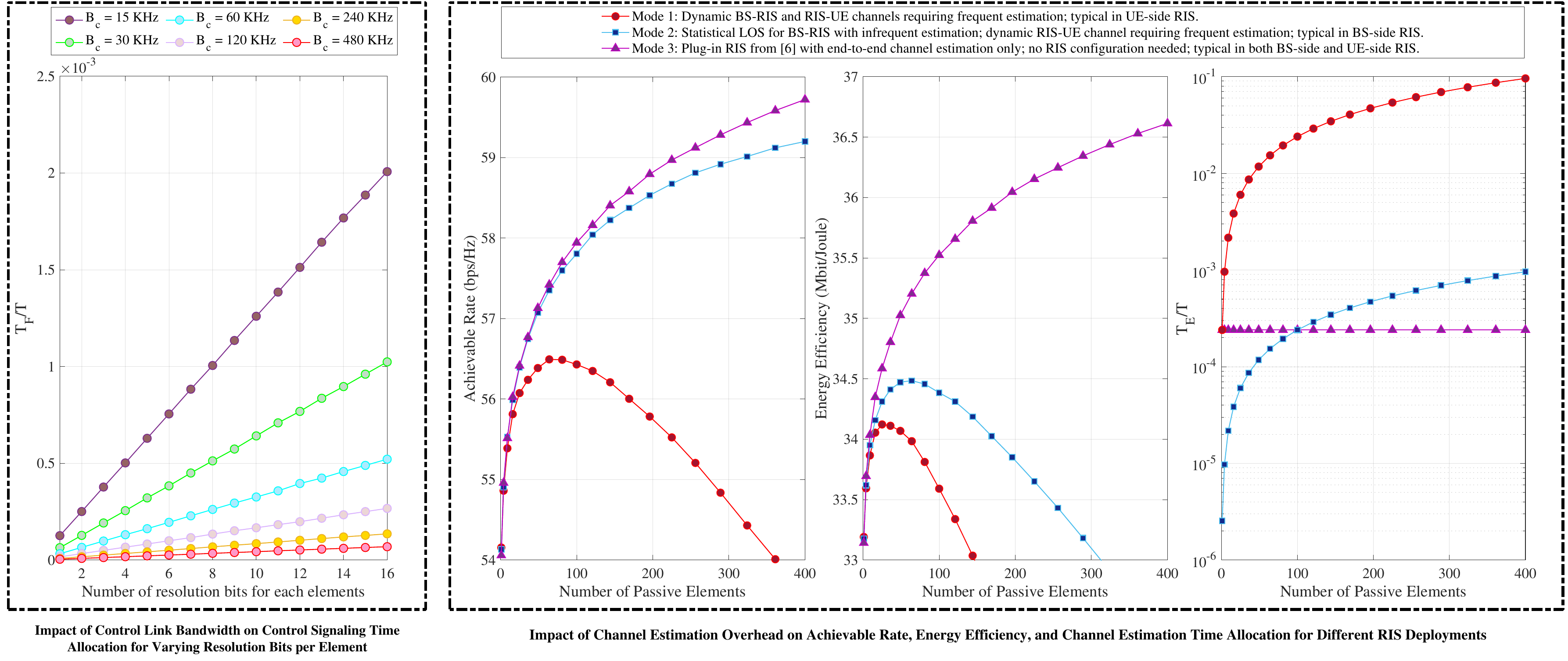}
    \caption{{Effects of control link bandwidth, control signaling, and channel estimation overhead on RIS performance metrics. The necessary formulas and parameters for these simulations can be found in \cite{9200578}.}}
    \label{fig: Sim}
\end{figure*}

\subsection{Phase Shift Profile Design}

One of the main and essential functions of RISs is their ability to focus the reflected signal into a specific spatial direction without using active RF components (passive beamforming). This is realized by adjusting the phase shifts of individual RIS elements relying on the CSI. 
Phase shift adjustment can be continuous or discrete. Continuous phase shifting results in superior performance enhancement, while discrete phase shift adjustment is cost-effective and less complex.

{Considering the availability of sophisticated hardware and the feasibility of implementing a high-capacity control link at the BS, configuring BS-side RIS with continuous (high-resolution) phase shifts is feasible, enabling optimal RIS tuning.}  
On the other hand, since the UEs typically suffer from limited {hardware}, and the UE-RIS control link cannot guarantee high-capacity communication, implementing low-complexity RIS configuration schemes in the UE-side RIS is essential{; accordingly, control signaling is preferred to be simple to prevent high signaling overhead and long latency. The first simulation in Fig. \ref{fig: Sim} shows that as control link bandwidth ($B_c$) increases, the ratio of control signaling time ($T_F$) to the total time block ($T$) decreases. This indicates that a high-capacity control link can transmit high-resolution control signals with minimal time allocation. The simulation assumes an RIS of size $10 \times 10$. Relevant parameters and formulas are given in \cite{9200578}. Besides,} heuristic approaches can be exploited to design a low-complex phase-shift configuration scheme \cite{10361836}. 
Another alternative can be adopting codebook-based schemes to decrease {overall} complexity and signaling overhead between the UE and RIS while maintaining satisfactory performance \cite{9952197}.

 
\section{Design Aspects for Peak Performance}

{

After optimizing the RIS tuning parameters, the focus must shift to key design considerations aimed at further mitigating multiplicative path loss and enhancing RIS beamforming gain. In this context, we examine two fundamental design parameters: RIS size and element type.

\subsection{RIS Size}
A key to mitigating the multiplicative path loss— the major challenge in RIS-based communication— is using an RIS with a large number of elements. However, deploying a large RIS requires sufficient physical space (e.g., a $10 \times 10$ planar RIS with half-wavelength spacing at $2.4$ GHz spans approximately $56 \text{ cm} \times 56 \text{ cm}$) and adequate resources to support channel estimation and the joint optimization of active BS/UE beamforming and passive RIS beamforming to ensure peak performance.

BS-side deployments generally offer ample space for larger RIS structures. Additionally, the advanced hardware at the BS can handle the computational complexity of large-scale RIS setups, including channel estimation and joint active-passive beamforming optimization. Furthermore, placing the RIS near the BS and leveraging the fixed positions of both allows for a stable statistical LOS channel, reducing the need for frequent BS-RIS channel estimation. This reduction in channel estimation overhead decreases time/power allocation to the pilot signals, significantly improving performance. 

Unlike BS-side RIS deployments, UE-side RIS implementations often face space constraints for large arrays. This limitation is evident in integrated RIS setups on the UE side. To address this, a multi-layer RIS structure was proposed in \cite{9598898}. Similarly, in O2I coverage extension scenarios where a transparent RIS is embedded in a window \cite{10056847}, space limitations can reduce the RIS size, potentially providing inadequate passive array gain to counteract multiplicative path loss.
However, exceptions exist; for example, deploying large RIS surfaces on building facades near dead zones, such as in street canyons, can provide substantial passive array gain to overcome multiplicative path loss. 
Additionally, in UE-side RIS designs, the dynamic channels demand frequent estimation. UEs also lack the advanced hardware and resources available in BS-side deployments, limiting optimal RIS operation. These limitations lead to poorer performance for UE-side RIS, compared to BS-side RIS. 

\begin{table*}[ht] 
\centering
\caption{{6G Requirements and RIS Deployment Types for BS-side and UE-side RIS}}
\renewcommand{\arraystretch}{1.5} 
\setlength{\tabcolsep}{6pt} 
\begin{tabular}{|m{2.2cm}|m{1.2cm}|m{1.7cm}|m{9.5cm}|m{1cm}|}  
\hline
\rowcolor{gray!20} 
{\textbf{6G Requirement}} & {\textbf{RIS Deployment}} & {\textbf{Recommended Strategy}} & {\textbf{Contributing Factors}} & {\textbf{Impact Level}} \\ \hline
\rowcolor{white}
{Latency} & {BS-side} & {Integrated}   & {Stable Control link; High Beamforming Gain} & {High} \\ \cline{2-2} \cline{4-5}
\rowcolor{white}
& {UE-side}      &  & {Beamforming Gain; Channel Estimation Overhead; Resource and Deployment Limitations} & {Medium} \\ \hline
\rowcolor{gray!10}
{Spectral Efficiency} & {BS-side} & {Integrated}   & {High Beamforming Gain} & {High} \\ \cline{2-2} \cline{4-5}
\rowcolor{gray!10}
& {UE-side}      & & {Beamforming Gain; Resource and Deployment Limitations} & {Medium} \\ \hline
\rowcolor{white}
{Energy Efficiency}   & {BS-side} & {Integrated}   & {Passive Element Utilization; Reduced Active Components} & {High} \\ \cline{2-2} \cline{4-5}
\rowcolor{white}
& {UE-side}      &  & {Passive Nature; Resource and Deployment Limitations} & {Medium} \\ \hline
\rowcolor{gray!10}
{User Experience Data Rate} & {BS-side} & {Integrated}   & {High Beamforming Gain; Robust Control} & {High} \\ \cline{2-2} \cline{4-5}
\rowcolor{gray!10}
& {UE-side}      &  & {Beamforming Gain; Compact and Efficient Designs; Resource and Deployment Limitations} & {Medium} \\ \hline
\rowcolor{white}
{Peak Data Rate}     & {BS-side}  & {Integrated}   & {High Beamforming Gain} & {High} \\ \cline{2-2} \cline{4-5}
\rowcolor{white}
 & {UE-side}      &  & {Beamforming Gain; Flexible Placement; Resource and Deployment Limitations} & {Medium} \\ \hline
\rowcolor{gray!10}
{Reliability}        & {BS-side}  & {Separated}  & {Limited Placement Flexibility} & {Low} \\ \cline{2-2} \cline{4-5}
\rowcolor{gray!10}
& {UE-side}      &  & {High Placement Flexibility} & {High} \\ \hline
\rowcolor{white}
{Connection Density} & {BS-side}  & {Separated}  & {Limited Placement Flexibility} & {Low} \\ \cline{2-2} \cline{4-5}
\rowcolor{white}
& {UE-side}      & & {High Placement Flexibility} & {High} \\ \hline
\end{tabular}
\label{Table}
\end{table*}

Fig. \ref{fig: Sim} shows how reduced channel estimation overhead affects channel estimation time $T_E$, achievable rate, and energy efficiency across RIS deployments. 
In order to have a fair comparison, we considered three different modes:
\begin{itemize}
    \item \textit{Mode 1}: Represents dynamic BS-RIS and RIS-UE channels that require frequent channel estimation, typical for \textit{UE-side RIS} deployments where both links are highly dynamic.

    \item \textit{Mode 2}: Assumes a statistical LOS for the BS-RIS channel with infrequent estimation and a dynamic RIS-UE channel needing frequent updates, typical for \textit{BS-side RIS} deployments.

    \item \textit{Mode 3}: Refers to the Plug-in RIS \cite{raeisi2023plug} with end-to-end channel estimation only, without requiring RIS configuration; applicable to both \textit{BS-side} and \textit{UE-side} deployments, offering minimal overhead.
\end{itemize}
In Mode 1, frequent channel estimation for both BS-RIS and RIS-UE links results in high overhead, reducing performance after a peak. Mode 2 benefits from a statistical LOS BS-RIS channel, requiring less frequent estimation, thus improving efficiency. Mode 3, with end-to-end estimation only, has minimal overhead, achieving the highest rate and efficiency consistently.

As shown in Fig. \ref{fig: Sim}, innovative designs like Plug-in RIS can overcome the overhead problem and improve performance. Developing such structures is crucial for addressing the practical challenges of RIS implementation.

\subsection{Element Type}
Deploying large RIS panels is not always feasible due to space constraints and the significant overhead load. As shown in Fig. \ref{fig: Sim}, increasing the number of RIS elements initially enhances the achievable rate and energy efficiency, but beyond a certain point, further increases lead to performance degradation due to the growing time and power demands for pilot signals. Thanks to the BS's advanced capabilities, BS-side RIS can maintain higher performance with lower resource allocation, as explained in the previous sub-section and demonstrated by Mode 2. In contrast, UE-side RIS faces more challenges due to dynamic channels from UE mobility and limited resources (shown via Mode 1 in Fig. \ref{fig: Sim}). In such cases, integrating active RIS elements can offer a solution by amplifying the reflected signal \cite{9998527}.
}

\begin{figure*}
    \centering
    \includegraphics[scale = 0.8]{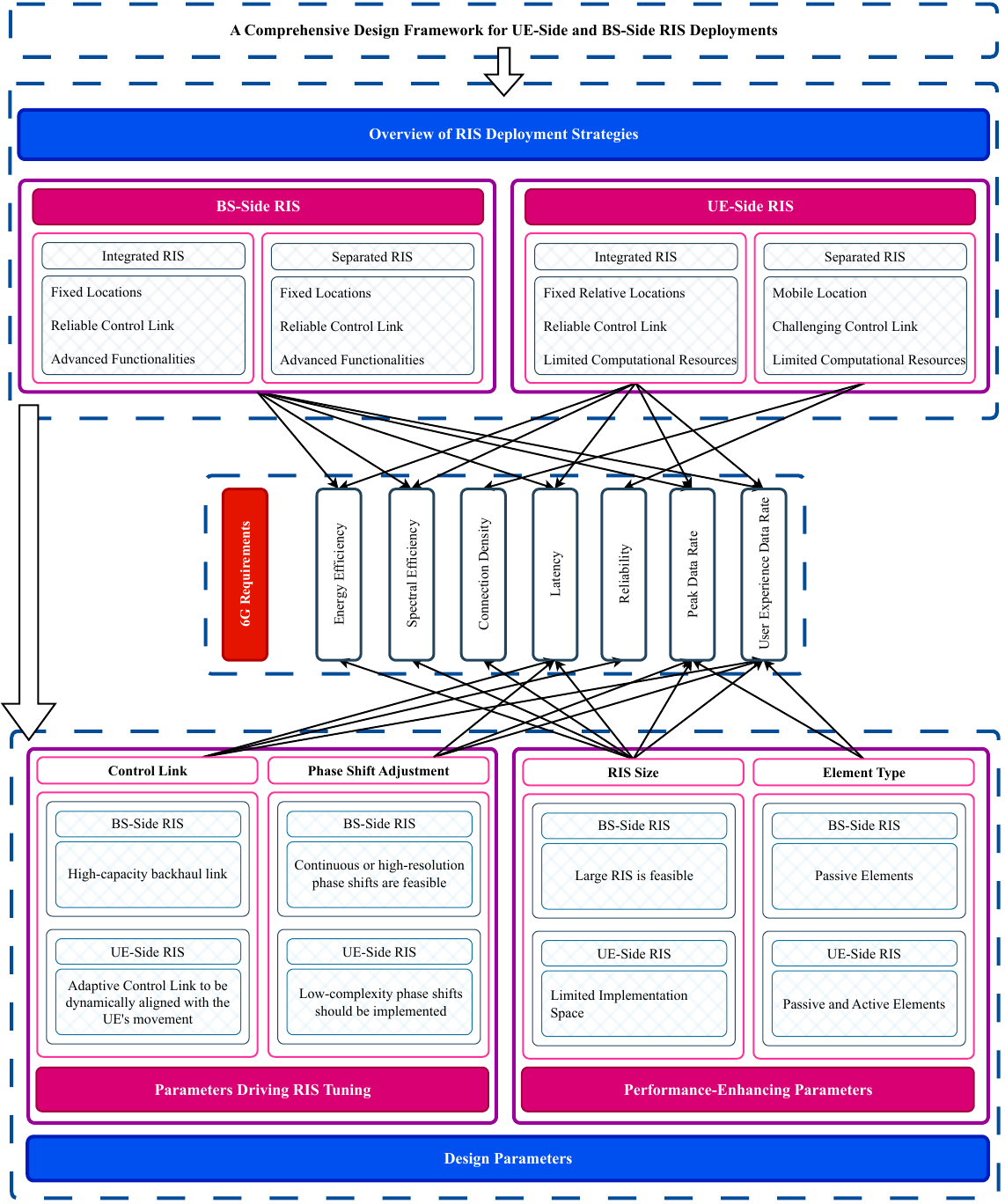}
    \caption{{Proposed design framework for BS-Side and UE-Side RIS deployments.}}
    \label{fig:Framework Overview}
\end{figure*}

\section{Advancing Towards 6G: How RISs Shape the Requirements of 6G Networks}
\label{Sec: 6G Requirements}

This section highlights different {features} associated with both BS-side and UE-side RIS deployments, with a particular emphasis on meeting the requirements of 6G networks. These requirements include high energy/spectral efficiency, low latency, massive connectivity, ultra-reliability, and high data rates. Specifically, we discuss how each RIS deployment can enhance various 6G requirements.

\subsection{Integrated or Separated RIS: Which Serves 6G Requirements Better?}\label{Sec: Separated/Integrated}

BS/UE-side integrated RIS results in the minimum path loss. Hence, we can maximize {data rate} and {spectral efficiency} while minimizing the communication {latency}. Furthermore, by integrating RIS into each terminal, we can replace active components with passive elements, resulting in enhanced {energy efficiency}. However, the BS/UE-side integrated design does not have flexibility in RIS placement; hence, it cannot circumvent blockages and provide redundant paths. Therefore, BS/UE-separated design is recommended to increase {connection density} and network {reliability}.

In conclusion, in a realistic network, both integrated and separated RIS designs should be adopted concurrently to meet the high standards of 6G networks.

\subsection{Towards 6G: The Strategic Role of BS-Side RIS}
Generally, BSs are placed within an environment to provide broad coverage. When an RIS is deployed near the BS, it inherits this wide coverage capability. This feature allows us to leverage the RIS as a passive beamformer, reducing costs, conserving power, and enhancing {energy efficiency} at the BS, as fewer active components like RF chains and phase shifters are required. 
{Furthermore, a large passive BS-side RIS can significantly increase beamforming gain, enhancing {user-experienced/peak data rate} and {spectral efficiency}, and reducing communication {latency}.}
As explained {earlier}, by placing RIS separated from the BS, the transmitted signal can circumvent the environmental blockages. However, to circumvent a faraway blockage, we may increase the BS-RIS distance, which increases multiplicative path loss. On the other hand, to keep the multiplicative path loss under a certain threshold, minimum BS-RIS distance should be maintained. In this point of view, BS-side RIS has minor flexibility in enhancing {connection density} and communication {reliability}.
Table \ref{Table} summarizes the influences of RIS on 6G requirements.

\subsection{UE-Side RIS: A Flexible Solution for 6G Network Requirements} 

Unlike BS, UEs generally do not have fixed positions and can move freely in the environment, potentially ending with dead zones. As explained in the last sub-section, BS-side RIS deployment might not effectively circumvent the blockages due to the limited placing flexibility {(see Scenario 3 in Fig. \ref{fig:General-Applications})}.
Therefore, as depicted in {the Scenario 4 in} Fig. \ref{fig:General-Applications}, the RIS can be positioned near the dead zone to circumvent the blockage more effectively while keeping the multiplicative path loss under a certain threshold. 
In this context, UE-side RIS can enhance {connection density} and communication {reliability} by reserving virtual LOS to the dead zone. 
On the other hand, integrating RIS within the UE realizes a large array at the UE \cite{9598898}, which can considerably enhance beamforming/array gain, directly increasing the {user-experienced/peak data rate} and {spectral efficiency} while reducing communication {latency}. 

\begin{figure*}
    \centering
    \includegraphics[width = \textwidth]{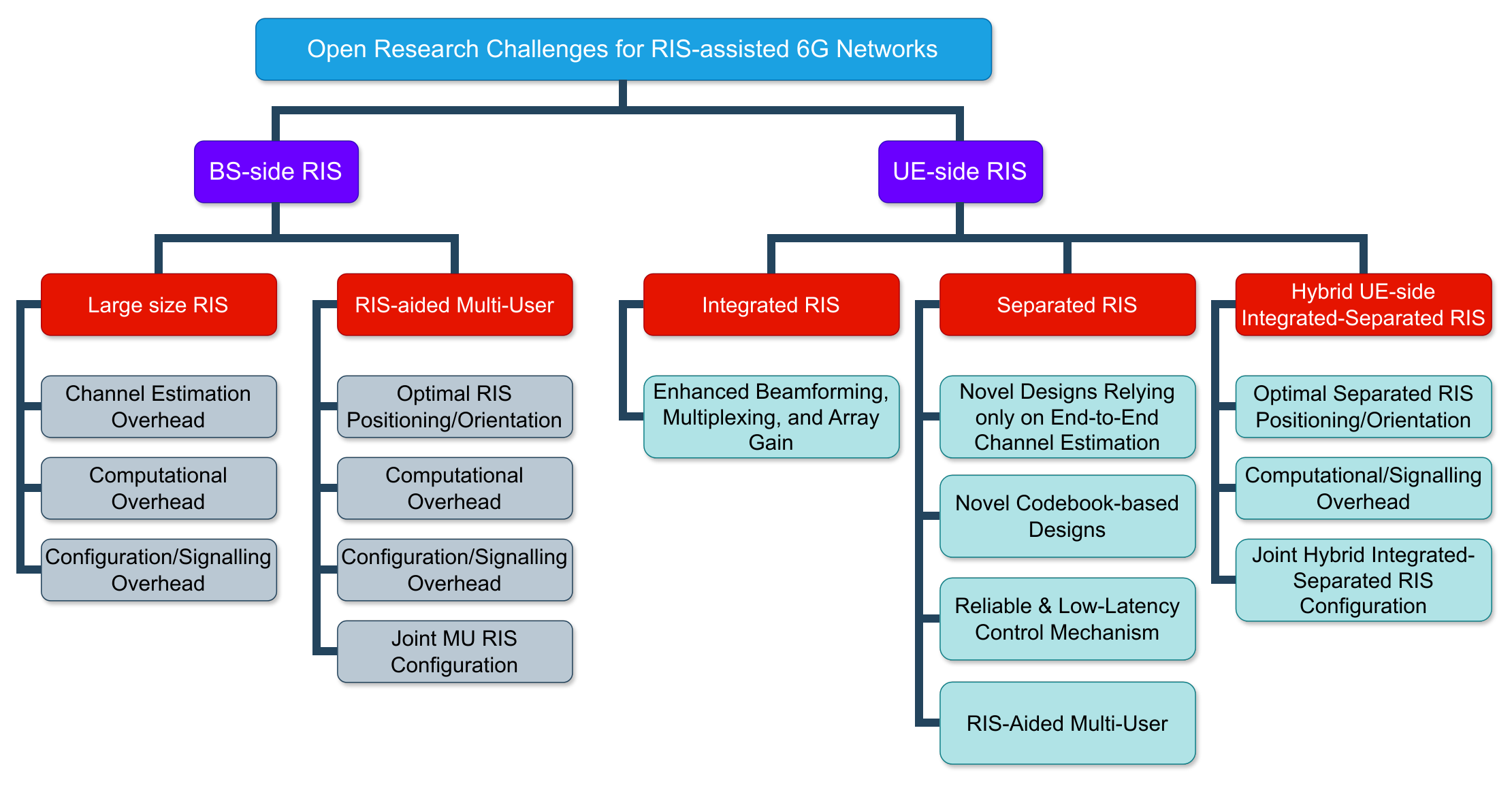}
    \caption{Open challenges for BS-side/UE-side RIS-aided 6G networks.}
    \label{fig:Open_Challenges}
    \vspace{-1.5em}
\end{figure*}

However, deploying large RIS near the UE{, either integrated into or separated from the UE,}  presents a unique challenge: the overhead associated with cascaded channel estimation and RIS configuration \cite{basar2021reconfigurable}. Given the UE's limited computational/power resources, these overheads can {increase} {system latency}{, as illustrated in Fig. \ref{fig: Sim}}. Accordingly, striking a balance between the RIS size and the signaling overhead while considering efficient channel estimation and RIS configuration strategies (like codebook-based configuration) is of great importance \cite{9200578}. Adopting novel designs like plug-in RIS \cite{raeisi2023plug}, which simplifies channel estimation and RIS configuration, can facilitate UE-side RIS deployment with higher {energy efficiency} and {lower latency}. {Fig. \ref{fig: Sim} demonstrates the superiority of Plug-in RIS over traditional RIS design.}

Table. \ref{Table} summarizes the effect of each RIS deployment on {6G} requirements. Generally, RIS positively impacts the 6G requirements; however, practical limitations should be considered for each deployment. For some 6G requirements, BS-side RIS provides higher performance enhancements, while UE-side RIS deployment is better for the other requirements. In a complete network, the hybrid implementation of these two deployments can complement the network needs.

\section{Conclusion and Future Horizon}

{RISs have demonstrated great potential in shaping intelligent radio environments for wireless communications tailored to meet the demands of future networks, such as ultra-reliable, low-latency communication and massive connectivity. To fully leverage RIS-assisted networks, it is essential to explore practical deployment strategies and their impacts.} {To this end, this article examined two practical RIS designs—BS-side and UE-side RIS—and proposed a framework with various design parameters.} Moreover, we have highlighted how these applications influence the requirements of 6G networks, showcasing their potential to revolutionize future communications. {Fig. \ref{fig:Framework Overview} outlines these RIS deployment strategies, highlighting key design parameters and their impact on 6G network performance.}

Despite significant progress, RIS technology faces practical challenges that need to be addressed for cost-effective and efficient integration into next-generation networks.
Fig. \ref{fig:Open_Challenges} outlines the future directions for BS/UE-side RIS deployments, highlighting key areas where innovation is crucial to unlock the full potential of RIS in future wireless communications.
{For the BS-side RIS, challenges include managing large-size RIS configurations, optimizing multi-user setups, and minimizing computational and signaling overhead. For the UE-side RIS, the challenges are categorized into integrated, separated, and hybrid configurations, emphasizing the need for enhanced beamforming, efficient channel estimation, and reliable control mechanisms. These challenges highlight the need for innovative designs and optimization strategies to maximize the benefits of RIS technology in future 6G networks.}

\bibliographystyle{ieeetr}
\bibliography{reference}

\end{document}